\documentclass[amssymb,prd,aps,showpacs,nofootinbib,preprint]{revtex4-1}
\usepackage{latexsym}
\usepackage{graphicx}
\usepackage{amsmath}
\usepackage[T1]{fontenc}
\usepackage[hang,small,up]{caption}
\makeindex
\begin{document}
\title{Pattern formation and coexistence domains for a nonlocal population dynamics}
\author{Jefferson A.R. da Cunha$^{3,4}$}\author{ André L.A. Penna$^{2,4}$}\author{Fernando A.
Oliveira$^{1,4}$}\email{fao@fis.unb.br}
\affiliation{Instituto de Física - Universidade de Brasilia, Brazil$^{1}$\\
FGA-Universidade de Brasília, Brazil$^{2}$\\
Instituto de Física Universidade Federal de Goiás, Goiânia, Brazil$^{3}$\\
International Center for Condensed Matter Physics\\CP 04455,
70919-970 Brasilia DF, Brazil,$^{4}$}


\begin{abstract}
\noindent
In this communication we propose a most general equation to study pattern formation for one-species population and their limit domains in systems of length $L$. To accomplish this we include non-locality in the growth and competition terms where the integral kernels are now depend on characteristic length parameters $\alpha$ and $\beta$. Therefore, we derived a parameter space $(\alpha,\beta)$ where it is possible to analyze a coexistence curve $\alpha^{*}=\alpha^{*}(\beta)$ which delimits domains for the existence (or not) of pattern formation in population dynamics systems. We show that this curve has an analogy with coexistence curve in classical thermodynamics and critical phenomena physics. We have successfully compared this model with experimental data for diffusion of {\em Escherichia coli} populations.
\end{abstract}
\date{\today }
\maketitle

{\bf Introduction -}
In recent years the phenomenon of pattern formation has been intensively studied to describe the spatial distribution of species in population dynamics. This amazing behavior of populations, observed under certain conditions, can be modeled by nonlinear equations of reaction and diffusion-type~\cite{Cross, Fife,Zhabotinsky,Legawiec,Vanag,Kepper,
Kawczynski}. Such mathematical models provide a rich structure to include
a variety of intra- or inter-specific interactions among species \cite{Anna,Fuentes}, as well as permitting to describe
many forms of effects of dispersal with or without memory effects \cite{Kenkre,Fuentes2,Showalter}.

On the other hand, the overwhelming majority of the studies have shown that Fickian-type diffusion \cite{Clerc,Bolster,Murray} is unable to describe spreading
of species in population equations formulated through reaction-diffusion models. Moreover, this is not the only
possible criticism that one can find at the ordinary nonlinear reaction-diffusion approach. In population dynamics
context, one believes that there is no real justification for assuming that interaction among species are, in fact, local.
There are many models in which such an assumptions become clearly unwarrantable, as for example, in competition of
one-species in a habitat where the system is rapidly equilibrated or in typical biological interactions where the
individuals intercommunicate via chemical means. Of course, these are typical nonlocal effects which should not be
overlooked. Other forms of nonlocal growth and interaction effects have been also observed when we deal with wide
variety of biological fields, such as epidemic spread in network \cite{Kenah,Barthelemy,Newman}, embryological development, and bacterial growth \cite{Mueller},
where the density of individuals involved are not small and the analysis of local or short-range diffusive flux is
not sufficient accurate to understand the dynamical aspects of these phenomena. In these cases we need to include the
contribution of long-range effects, and then to analyze the domains limits for the existence of patterns in these systems.

The present communication is an attempt to build a most general equation to study effects of pattern formation in one-species population dynamics. Our starting point is to write an equation which includes nonlocal growth and interaction terms involving long-range effects in the system. This equation can be written as
\begin{eqnarray}
\label{interaction_competition}
\frac{\partial u(x,t)}{\partial t}=a\int_{\Omega}g_{\alpha}(x-x')u(x',t)dx'-bu(x,t)\int_{\Omega}f_{\beta}(x-x')
u(x',t)dx'\,\,,
\end{eqnarray}
where $u(x,t)$ describes the population density with growth $a$
and competition $b$ terms. Then we denote by $g_{\alpha}(x-x')$ the correlation growth function, which weights the
growth of a population in the domain $\Omega$ for a specific growth length parameter $\alpha$. We call $f_{\beta}(x-x')$
the correlation competition function which weights the interaction among the constituents of the population for a competition length parameter
$\beta$ in the domain $\Omega$. In the intuit to modeling population dynamics, we have assumed that the kernels are symmetric functions, such that $f_{\beta}\rightarrow 0$ and $g_{\alpha}\rightarrow 0$ as $|x-x'|\rightarrow\infty$. An of advantages of Eq. (\ref{interaction_competition}) is that it provides an useful concept to describe a great variety of long-range diffusive
effects in physical systems, by permitting we can absorb all higher derivatives (diffusion and dissipation terms) in only integral term.
Moreover this model is then parametrized mainly by $\alpha$ and $\beta$ quantities, and this allows us
to suppose on the existence of values ($\alpha,\beta$) for which there are pattern formation. Indeed the relation between
these length domains offers a simplest model to deal with quantitative estimates of experimental
data related to the growth dynamics of bacteria, for example. In this case, specific domains which show the existence (or not) of patter formation, which incorporates long-range effects as well, is important for the physical description of spreading of individuals with nonlocal dynamics.

Starting from Eq. (\ref{interaction_competition}) we can derive important connections with classical population models by carrying
out appropriated limits. For example, if $f_{\beta}(x-x')=g_{\alpha}(x-x')=\delta(x-x')$ we get the logistic
equation \cite{Murray}
\begin{equation}
\frac{\partial u(x,t)}{\partial t}=au(x,t)-bu(x,t)^2.
\end{equation}
One can also consider $g_{\alpha}(x-x')$ with a finite range, such that we can expand the growth term as
\begin{equation}
a\int_{\Omega}g_{\alpha}(x-x')u(x',t)dx'=\sum_{m=0}^{\infty}\frac{a\overline{y^{2m}}}{(2m)!}
\frac{\partial^{2m}}{\partial x^{2m}}u(x,t)\,,
\label{gener}
\end{equation}
where $y=x-x'$ and the k-moments are
\begin{equation}
\label{med}
\overline{y^{k}}=\int y^{k}g_{\alpha}(y)dy\,.
\end{equation}
Using the above procedure with $f_{\beta}(y)=\delta(y)$ and retaining the first two terms in Eq. (\ref{gener})
we get the ordinary Fisher equation
\begin{equation}
\frac{\partial u(x,t)}{\partial t}=D\frac{\partial^2u(x,t)}{\partial
x^2}+au(x,t)-bu^2(x,t).
\label{eq.Fisher}
\end{equation}
Here we show a very important point: the first gain with the nonlocal growth term is the possibility to connect the
growth rate $a$ with the diffusion constant
\begin{equation}
D=\frac{a\overline{y^{2}}}{2}\,.
\label{Y_momentum}
\end{equation}
Note that this equation shows that a species with a large growth rate has "more need" for diffuse behavior, {\it i.e.} a large
growth rate creates a large pressure proportional to the concentration gradient $D\frac{\partial}{\partial x}u(x,t)$ which increases
the diffusion. It shows that the diffusion is intrinsically related to the existence of the nonlocal growth term $g_{\alpha}$, {\it i.e.} it is necessary to exist a second moment $\overline{y^{2}}\neq 0$ in according to Eq. (\ref{med}).

The higher order terms ($m>1$) in expansion of Eq. (\ref{gener}) yield the dispersive terms. It is interesting
to note that if $g_{\alpha}$ is not even  the first derivative yields the convective term
$v\frac{\partial}{\partial x}u(x,t)$, where we can obtain the convective velocity as $v=\overline{y}a$. On the other hand, if we have an asymmetric $g_{\alpha}$ it corresponds to a convective drift \cite{Jefferson}. If we keep the expansion up to second order and $f_{\beta}(y)$ in the second integral, we get nonlocal Fisher equation
\begin{equation}
\frac{\partial u(x,t)}{\partial t}=au(x,t)+D\frac{\partial^{2}u(x,t)}{\partial x^{2}}-bu(x,t)\int_{\Omega}f(x-x')u(x,t)dx\,\,,
\label{eq.Fisher2}
\end{equation}
which has been widely used by many authors to discuss pattern \cite{Fuentes}. Note that is Eq. (\ref{interaction_competition}) is most general, incorporating all the previous equations.

{\bf Perturbative analysis -}
In the study of pattern formation through a model of population dynamics it is usual to calculate a quantity known as the growth rate of pattern $\gamma$ \cite{Fuentes, Kenkre} which leads to the pattern formation in the system. Therefore, we first
shall start with the perturbative analysis through the function
\begin{equation}
u(x,t)=\frac{a}{b}+\epsilon \exp{\Big(ikx+\phi(k)t\Big)},
\label{test_function}
\end{equation}
where $a/b$ is the homogeneous steady state solution,
constant in space and time. The term $\epsilon\exp(ikx)\exp(\phi t)$
is a perturbation to the steady state that will grow or die out,
depending on the values of the
wavenumbers $k$.
Substituting Eq.~(\ref{test_function}) into Eq.~(\ref{interaction_competition})
and retaining only first order perturbative terms, we find a
dispersion relation between the complex pattern growth rate $\phi$ and the wavenumber $k$, given by
\begin{eqnarray}
&&\phi(k)=a\Big({\cal F}_{c}\{g_{\alpha}(y)\}-{\cal F}_{c}\{f_{\beta}(y)\}\Big)+
ia\Big({\cal F}_{s}\{g_{\alpha}(y)\}-{\cal F}_{s}\{f_{\beta}(y)\}\Big)\,\,,
\label{dispersion_relation}
\end{eqnarray}
where ${\cal F}_{c}\{\cdot\}$ and ${\cal F}_{s}\{\cdot\}$ are, respectively, the Fourier cosine and sine transform
of the influence function $g_{\alpha}(y)$ and $f_{\beta}(y)$ (assumed to be even). Therefore,
only the real part of the complex growth rate $\phi(k)=\gamma(k)+i\xi(k)$, where $\gamma(k)=a\Big(\int_\Omega
g_{\alpha}(y)cos(ky)dy-\int_\Omega
f_{\beta}(y)cos(ky)dy-1\Big)$,
{\em i.e.} the growth rate of pattern $\gamma(k)$, will be important to determine whether the perturbation with wavenumber $k$
will die out or will generate a pattern, for negative or positive values, respectively.
Now let us consider the simple case of the square interaction influence function given by
\begin{equation}
\label{influence_function}
f_{\beta}(y)=\frac{1}{2\beta}\big[\Theta(\beta-y)\Theta(\beta+y)\big]\,,
\end{equation}
where $\Theta$ refers to the Heaviside function and $\beta$ is the cut-off range
($0<\beta<L$, where $L$ is the size of the system). If we consider similar relation for $g_{\alpha}$ with cut-off
$0<\alpha<L$, from the condition (\ref{influence_function}), $\gamma(k)$ is given by
\begin{equation}
\gamma(k)=a\Big(\frac{\sin(k\alpha)}{k\alpha}-\frac{\sin(k\beta)}{k\beta}-1\Big).
\label{rate_alpha_beta}
\end{equation}
Therefore we can study self-organization of the equation (\ref{interaction_competition}) considering that the system
depends physically on the domain of the functions $g_{\alpha}$ and $f_{\beta}$. In this case, pattern formation
appears when wave numbers $k$, in growth rate of pattern, obey the condition $\gamma(k)>0$. Note that for $\alpha=0$,
$\gamma(k)$ is larger. This will happen for lower diffusive systems $D\approx 0$.

\begin{figure}[h!]
\begin{centering}
\rotatebox{0}{\resizebox{8.0cm}{!}{\includegraphics{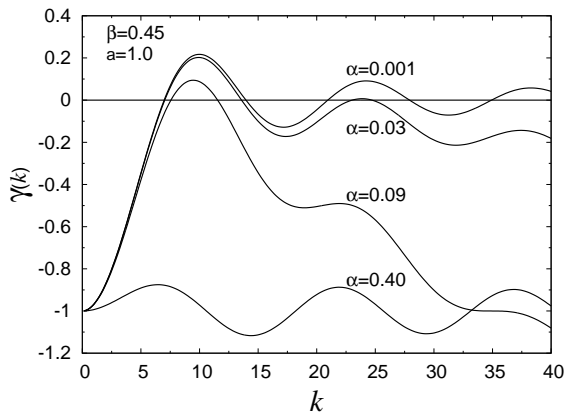}}}
\caption{The real part of growth exponent $\gamma(k)$ as a function of $k$ plotted for different values of the
correlation length of growth $\alpha$ with a length interaction of individuals $\beta$ fixed. The pattern formation
appear for those values of $k$ for which $\gamma$ is positive.}
\label{fig.rate}
\end{centering}
\end{figure}

The Eq. (\ref{rate_alpha_beta}) is plotted in Fig.
\ref{fig.rate} for different values of the growth length $\alpha =(0.01, 0.03, 0.09, 0.40)$ with
competition length parameter $\beta=0.45$ fixed. In this figure, we verified that when $\alpha < \beta $ we may have $\gamma(k)>0$. If $\alpha\rightarrow\beta$, the function $\gamma(k)$ becomes negative. This
behavior of $\gamma(k)$ is very important to determine if we have a large or a negligible amplitude of pattern.
We show in Fig. \ref{fig.rate} that pattern formation appears for values $\alpha<\beta$ with $\gamma(k)>0$. This
behavior is also verified later with numerical results, see Fig. 4, and discussed through experimental values.

{\bf Numerical method -}
To solve Eq.~(\ref{interaction_competition}) numerically, we applied the Operator Splitting Method
(OSM)~\cite{William}. By this method, the operator of the differential equation is split into several parts,
which act additively on $u(x,t)$. If we write Eq.~(\ref{interaction_competition}) as
\begin{equation}
\frac{\partial u(x,t)}{\partial t}=\hat{T} u(x,t),
\label{eq15}
\end{equation}
where $\hat{T}$ is the total operator, then
\begin{equation}
\hat{T} u(x,t)=\hat{T}_{grow}u(x,t)+\hat{T}_{int}u(x,t), \label{eq16}
\end{equation}
with
\begin{eqnarray}
&& \hat{T}_{grow}u(x,t)=a\int_{\Omega}g_{\alpha}(x-x')u(x',t)dx'\\
\nonumber\\
&& \hat{T}_{int}u(x,t)=-b u(x,t)\int_{\Omega}f_{\beta}(x-x')u(x',t)dx'.
\label{eq17}
\end{eqnarray}
In the latter equations $\hat{T}_{grow}$ and
$\hat{T}_{int}$ are nonlocal growth and nonlocal interaction
operators, respectively. In our numerical calculations, we have used periodic boundary conditions
$u(x=0,t)=u(x=L,t)$ with spatial period $L$. For each part of the operator, we apply a known difference scheme for
updating the function $u(x,t)$ from step $j$ to step $j+1$.

\begin{figure}[h!]
\begin{centering}
\rotatebox{-90}{\resizebox{7.0cm}{!}{\includegraphics{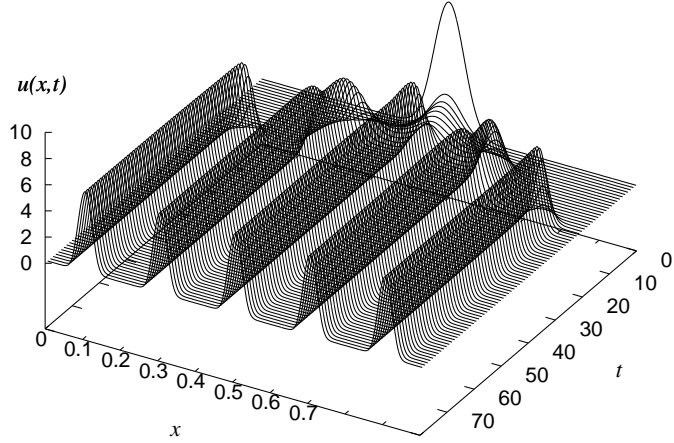}}}
\caption{The typical pattern formation on density $u(x,t)$ as a function of $x$ and $t$ in arbitrary units. The
growth rate and interaction rate are $a=b=1.0$. The competition length parameter $\beta=0.15$ and the growth
length parameter $\alpha=0.009$.}
\label{3d_sdeady_state}
\end{centering}
\end{figure}

\begin{figure}[h!]
\begin{centering}
\rotatebox{270}{\resizebox{10.0cm}{!}{\includegraphics{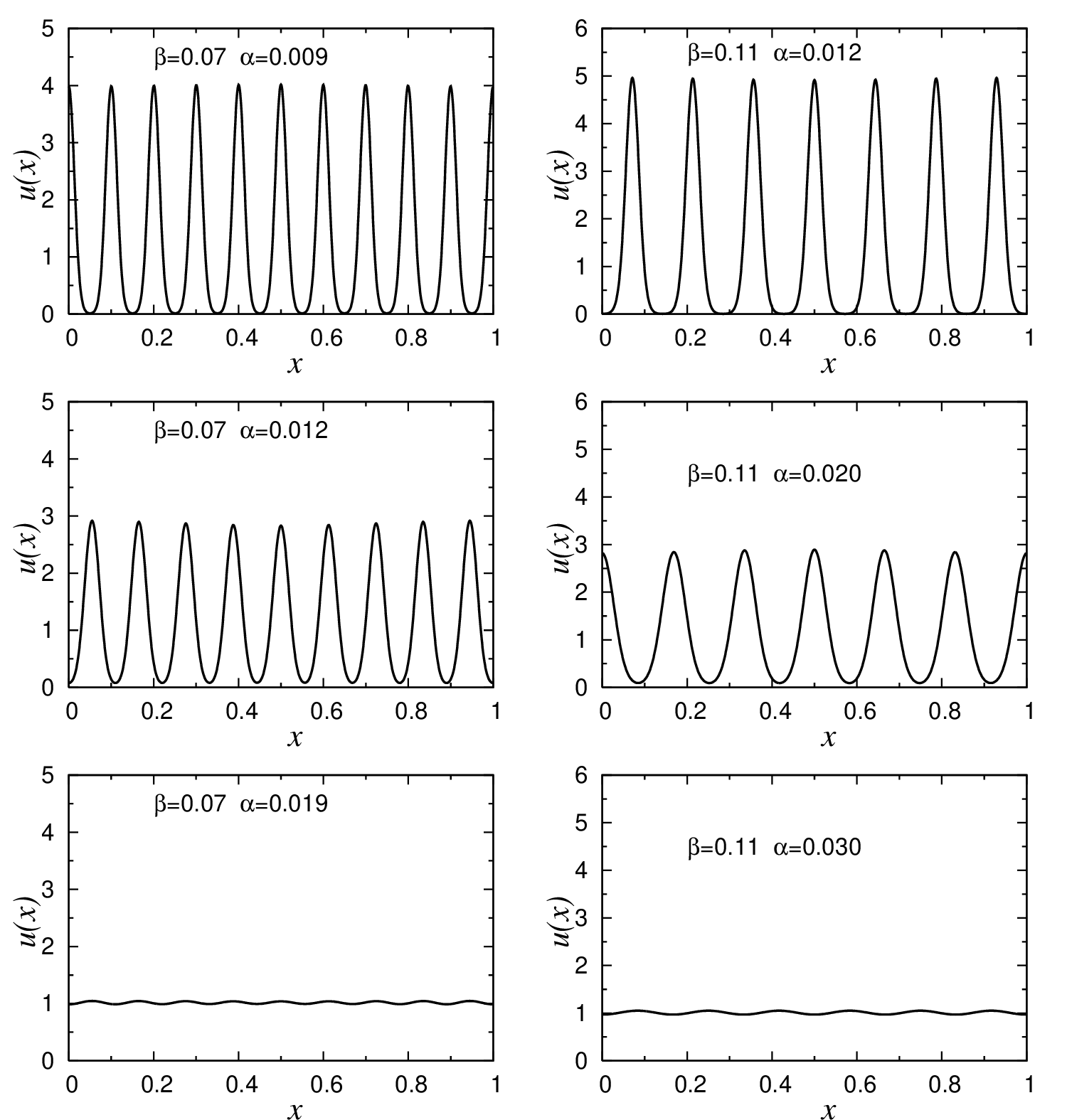}}}
\caption{Snapshots of the stationary state $u(x)$ for some values of competition length parameter $\beta$ and
growth length parameter $\alpha$ with correlation competition function and correlation growth function
Eq. (\ref{influence_function}). In this snapshots we consider the growth rate $a$ and competition rate $b$
as $1.0$. For a fixed $\beta$ as $\alpha$ increases
the pattern disappears.}
\label{snapshots_stationary_state}
\end{centering}
\end{figure}

\begin{figure}[h!]
\begin{centering}
\rotatebox{0}{\resizebox{8.0cm}{!}{\includegraphics{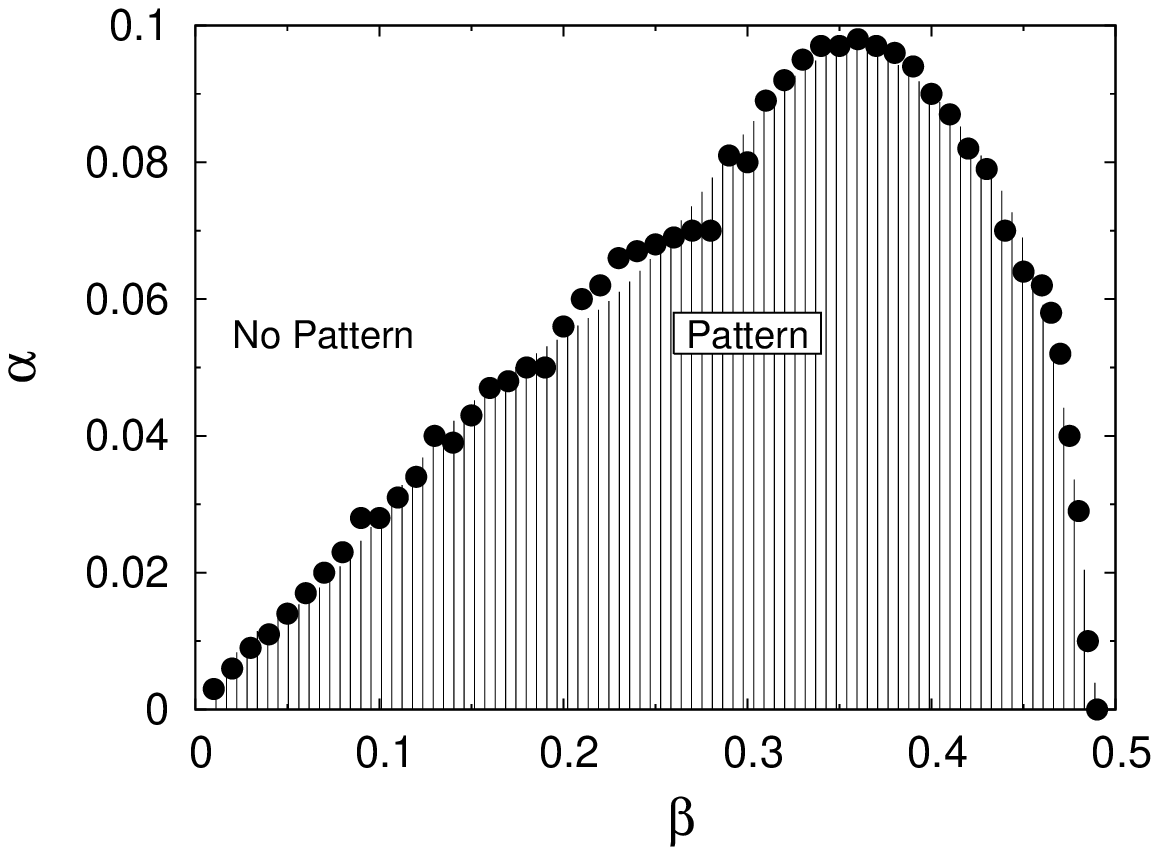}}}
\caption{The phase diagram of critical correlation growth length $\alpha_{c}$ as a function of critical
competition length parameter $\beta_{c}$. The Pattern and Non Pattern region indicate the separation of
large-amplitude patterns and negligible-amplitude patterns for Eq.(\ref{density_initial_distribution}).}
\label{fig_beta_alpha}
\end{centering}
\end{figure}

In Fig. \ref{3d_sdeady_state} we show the evolution of $u(x,t)$. We start with a distribution of individuals
\begin{equation}
u(x,0)=\frac{1}{\Gamma}\exp\left[-{\frac{(x-x_{0})^2}{2\sigma^2}}\right],
\label{density_initial_distribution}
\end{equation}
where
$\Gamma=\sqrt{\frac{\pi}{2}}\sigma\left[
\text{erf}\left(\frac{x_{0}}{\sqrt{2}\sigma}\right)+\text{erf}\left(\frac{L-x_{0}}{\sqrt{2}\sigma}\right)\right]
$, and we see the evolution to a state which exhibits pattern. We use $\sigma=0.3$, $x_{0}=0.5$ and $L=1.0$. The spacial and time
increments are $\delta x=1\times 10^{-3}$ and $\delta t=1\times 10^{-2}$. In all simulations we use $a=b=1.0$.
Several numerical experiments show the final state independent of the initial conditions \cite{Jefferson}. These
simulations are fundamental to show the pattern formation that appears after a long time for bacterial growth
\cite{Anna,Fuentes,Fuentes2,Kenkre,Perry}. Similar simulations are used to compose Fig.
\ref{snapshots_stationary_state} and Fig. \ref{fig_beta_alpha}.

In Fig. \ref{snapshots_stationary_state}, we show the evolution after $60000$ time steps where the density has reached its final form. Each curve is similar to the simulations described in Fig. \ref{3d_sdeady_state}. The first column is for $\beta=0.07$ and for up to down
$\alpha=(0.009, 0.012, 0.019)$. The second column has $\beta=0.11$ and for up to down
$\alpha=(0.012, 0.020, 0.030)$. For each $\beta$ the lower curve represents the hight value of $\alpha$, for which there is no more pattern formation, {\it i.e.} for $\alpha \geq \alpha^{*}$, where we get $u(x,t)=a/b$.

In Fig. \ref{fig_beta_alpha}, we show the region in the space $(\alpha, \beta)$ where pattern can exist. For each
$\beta$ the points represent the $\alpha^{*}$ above which there is not more pattern, such as described in Fig.
\ref{snapshots_stationary_state}. The shadow area is limited by the coexistence curve
\begin{equation}
\label{field2}
\alpha^{*}(\beta)=P(\beta)(\beta_{c}-\beta)^{\mu},
\end{equation}
which is the best fit of the points. $P(\beta)$ is a polynomial with no roots in the region $0<\beta<\beta_{c}$.
For $\beta\rightarrow\beta_{c}$ we get from Eq. (\ref{influence_function}) $\beta_{c}=1/2$, {\it i.e} the function $f_{\beta}(y)$ weights equally the all the space $0<x<L$, and consequently we have no pattern formation for $\beta \geq \beta_{c}$. Finally, from the data we have estimated the value $\mu=0.53\pm 0.06$ at the vicinity of $\beta_{c}$ for the exponent in Eq. (\ref{field2}).

{\bf Experimental data -} Starting from Eq.
(\ref{influence_function}) we can compute $\overline{y^{2}}=\alpha/3$. By inserting this result into Eq. (\ref{Y_momentum}) we get
\begin{equation}
\alpha=\sqrt{\frac{6D}{a}}.
\label{alpha_function_difusion}
\end{equation}
Now using the experimental values for $a=(2.23\pm 0.2)\times 10^{-4}s^{-1}$ and
$D=(2.2\pm 0.2)\times 10^{-5}cm^{2} s^{-1}$ obtained by N. Perry \cite{Perry}, for systems with {\em Escherichia coli}
populations, we can estimate the value of $\alpha$, given by $\alpha=(7.70\pm 0.09)mm$. Then to form pattern in
a finite system of length $L$, $\beta$ must be inside the shadow area of Fig. 4. In fact, the value of $\alpha$ as obtained in
Eq.(\ref{alpha_function_difusion}) is determined by the coefficient of diffusion $D$ of system and it establishes a
lower referential limit for the presence (or not) of pattern formation. For $\beta \ll L$, we are in the linear
part of Eq. (\ref{field2}) and we get $\alpha=0.3\beta$, therefore we have pattern for $\beta >25.4mm$.
In this case, only experimental values of width of influence function $\beta >25.4mm$ permit patterns,
which are in concordance with our theoretical and numerical results. Moreover our formulation allow us to
analyze pattern formation as an interplay between two length parameters $\alpha$ and $\beta$. It is important to note that the fact that $0 <\alpha < \beta$ is not just a curiosity of the theory, it is one of its major result. Without a finite value of  $\alpha$, there will be no diffusion, which is fundamental for any species, and so reproduction and propagation are associated. Consequently one should expect a non null $\alpha$. On the other hand if $\alpha$ is too large, bonds are tight, and they may face difficult to meet and consequently to reproduce. This phenomena described here for pattern formation in bacterial colony can be observed in large animals with migrate habits, such as deer and woolf, they travel with the only family, we call this phenomena \emph{the faithful sailor travel}.

{\bf Conclusion-}
The presence of memory, non-locality in time,  have been  used to explain ergodicity violation in particle diffusion \cite{Lapas08,Burov10}. Since pattern formation implies in ergodicity breaking, one could expect that a nonlocal space kernel would yield that. Consequently, we proposed here a new  formulation for population dynamics, which includes a growth and a competitive nonlocal terms. The presence of two kernels $g_\alpha(x)$ and $f_\beta(x)$ demand the existence of a growth length parameter $\alpha$ and of a competition parameter $\beta$.  Particular values for the kernels yield most of the previews formulations of population dynamics. We obtain a domain region $0< \alpha < \beta$ where patterns may arise, a coexistence curve similar to those in phase transition, and a direct connection between the diffusion constant $D$ the growth rate $a$ and the mean square deviation $\overline{y^2}=\int f_\alpha (y)y^2 dy$ which is a function of $\alpha$. More results can be obtained from this formulation, however there are some restrictions, we need more detailed dynamical experiments in growth, in such way that we can propose more elaborated kernels. Those present here, Eq (4), gives us a rough idea of the dynamics. More accurate  $g_\alpha(x)$ and $f_\beta(x)$ will permit us to get a better description and a generalization for two dimensions.

\end{document}